\documentclass[12pt]{article}

\usepackage{graphicx,amssymb,amsmath,empheq,color}
\usepackage{simpler-wick}

\usepackage{hyperref}
\hypersetup{colorlinks = true, linkcolor=black, citecolor=black, urlcolor=blue}
\usepackage{url}

\makeatletter

\newlength{\fighskip} \fighskip=2pt
\newlength{\figvskip} \figvskip=3pt

\newcommand*{\figbox}[2]{{
  \def\figscale{#1}
  \def\arraystretch{0.8}
  \arraycolsep=0pt
  \begin{array}{c}
    \vbox{\vskip\figscale\figvskip
      \hbox{\hskip\figscale\fighskip
        \includegraphics[scale=\figscale]{#2}}}
  \end{array}}}

\newcommand*{\widebox}[1]{\setlength{\fboxsep}{1ex}%
  \fbox{#1}}
\newcommand*{\wideboxed}[1]{\setlength{\fboxsep}{1ex}%
  \fbox{\m@th$\displaystyle#1$}}

\def\ubrace#1_#2{%
  \underbrace{#1}_{\hb@xt@\z@{\hss$\scriptstyle#2$\hss}}}

\newcommand{\blangle}{\bigl\langle}
\newcommand{\brangle}{\bigr\rangle}
\newcommand{\dlangle}{\langle\kern-1.5pt\langle}
\newcommand{\drangle}{\rangle\kern-1.5pt\rangle}
\newcommand{\bdlangle}{\blangle\kern-3pt\blangle}
\newcommand{\bdrangle}{\brangle\kern-3pt\brangle}

\newcommand*{\bra}[1]{\langle#1|}
\newcommand*{\ket}[1]{|#1\rangle}

\newcommand*{\bbra}[1]{\blangle#1\big|}
\newcommand*{\bket}[1]{\big|#1\brangle}

\newcommand*{\corr}[1]{\langle{#1}\rangle}
\newcommand*{\bcorr}[1]{\blangle{#1}\brangle}

\newcommand{\kap}{\varkappa}

\newcommand{\calH}{\mathcal{H}}

\newcommand{\R}{\mathrm{R}}
\newcommand{\A}{\mathrm{A}}
\newcommand{\Corr}{\mathcal{B}}
\newcommand{\bath}{\mathrm{b}}
\newcommand{\sys}{\mathrm{B}}

\newcommand{\TFD}{\mathrm{TFD}}
\newcommand{\tG}{\tilde{G}}
\newcommand{\UV}{\mathrm{UV}}

\newcommand{\bydef}{:=}
\newcommand{\wh}{\widehat}

\newcommand*{\trans}{{\mathpalette\@trans{}}}
\newcommand*{\@trans}[2]{\raisebox{\depth}{$\m@th#1\intercal$}}

\DeclareMathOperator{\Tr}{Tr}

\DeclareMathOperator{\TT}{\mathbf{T}}
\DeclareMathOperator{\tTT}{\widetilde{\mathbf{T}}}
\DeclareMathOperator{\BB}{\mathbf{B}}

\definecolor{DarkGreen}{rgb}{0.0, 0.5, 0.0}

\makeatother

\title{Perturbative calculations of entanglement entropy}

\author{Pouria Dadras\footnote{pdadras@caltech.edu},
Alexei Kitaev\footnote{kitaev@caltech.edu},
and Pengfei Zhang\footnote{pengfeizhang.physics@gmail.com}\\
\normalsize\it California Institute of Technology, Pasadena, CA 91125, U.S.A.\vspace{0.5cm}}

\date{October 25, 2022}

\begin{document}

\setcounter{tocdepth}{2}

\maketitle
\begin{abstract}
This paper is an attempt to extend the recent understanding of the Page curve for evaporating black holes to more general systems coupled to a heat bath. Although calculating the von Neumann entropy by the replica trick is usually a challenge, we have identified two solvable cases. For the initial section of the Page curve, we sum up the perturbation series in the system-bath coupling $\kappa$; the most interesting contribution is of order $2s$, where $s$ is the number of replicas. For the saturated regime, we consider the effect of an external impulse on the entropy at a later time and relate it to OTOCs. A significant simplification occurs in the maximal chaos case such that the effect may be interpreted in terms of an intermediate object, analogous to the branching surface of a replica wormhole.
\end{abstract}

\enlargethispage{10pt}
\tableofcontents
\newpage

\section{Introduction}\label{sec:introduction}

While not an observable quantity, entropy is useful as an abstract measure of active degrees of freedom and correlations in the system. Quantum correlations can be elusive, particularly in black holes, where the classical space-time picture is incomplete. There has been a long but ultimately successful chase of correlations in the Hawking radiation. If a black hole forms from an object in a pure quantum state and then evaporates, the resulting radiation must also be in a pure state. Thus, it is strongly (albeit nonlocally) correlated. The general form of such correlations was predicted by Don Page~\cite{Page93}, who considered a black hole as a generic quantum system. Still, it long remained unclear how such correlations could emerge in semiclassical gravity. Some important works that contributed to the solution include the Dray-'t\,Hooft mechanism of gravitational interaction between infalling matter and subsequent radiation~\cite{DtH85,tH90} and the calculation of out-of-time-order correlators (OTOCs) in the black hole setting~\cite{ShSt14}. However, the OTOC physics is relevant on short time scales and explains correlations that are present not in the radiation itself but relative to a purifying system~\cite{HaPr07} (that is, under the assumption that the black hole is part of a thermofield double and that we have unrestricted access to the other part). The recent breakthrough in understanding the correlations developing over the Page time~\cite{AHMST19,PSSY19} required a careful formulation of the problem, which we will now summarize.

The problem at hand is a semiclassical one. We do not have a complete theory of quantum gravity, nor should it be required. When working at the semiclassical level, it is not possible to derive long-term evolution from the short-term one. Rather, one should look for a global solution, which may depend on the quantity of interest. We consider the entanglement entropy between the black hole and the emitted radiation at a particular time $t$. So let $\rho=\rho(t)$ be the black hole's density matrix; we want to compute its von Neumann entropy, $S(\rho)=-\Tr(\rho\ln\rho)$. The latter is expressed as the $s\to 1$ limit of the $s$-Renyi entropy,
\begin{equation}
S_s(\rho)=\frac{1}{1-s}\ln\Tr\rho^s.
\end{equation}
For integer $s$, the expression $\Tr\rho^s$ may be interpreted as the partition functions of $s$ replicas of the system. 

Now, it turns out that the transition from the early phase of the black hole evaporation (when the radiation is uncorrelated as the naive theory predicts) to the later phase (when the entanglement entropy equals the black hole's coarse-grained entropy) is first order. The later phase is described by a new type of space-time geometry, the replica wormhole~\cite{AHMST19,PSSY19}. Although choosing the correct solution of the two is a global problem, each of them can be examined locally. We will study some properties of both solutions for general many-body systems, where the geometric description is not applicable.

The von Neumann and Renyi entropies are nonlinear functions of the quantum state, which is why they are not observables. However, the logarithmic nonlinearity is mild, such that in the thermodynamic limit, $S(\rho)$ is determined by typical microstates that contribute to the mixed state $\rho$. In contrast, Renyi entropies are often dominated by a fraction of microstates of tiny overall weight. This distinction is also evident from the replica wormhole picture. The $s$-Renyi entropy is related to an $s$-fold cover of space-time, whose metric is different from the physical one. But when we analytically continue the solution in $s$ and take $s$ to $1$, we get the standard metric with an additional piece of data, the branching surface. Thus, the $s\to 1$ limit is essential for compatibility with the usual (non-entropic) physics. Our main technical advance is how to take this limit in some specific cases.

\section{Early phase of entanglement growth}

We adopt a simpler variant of the evaporation problem, where instead of radiating energy, the system comes into contact with a heat bath at the same temperature. Turning the system-bath interaction on represents a slight change in the Hamiltonian and results in a brief period of non-equilibrium dynamics. Then a steady state is achieved such that all simple correlation functions are thermal. However, if the system's initial state was pure (though mimicking the thermal state), its von Neumann entropy will grow at a constant rate. We focus on this regime as well as the very beginning of quantum evolution. The entropy growth eventually saturates at the thermal (i.e.\ coarse-grained) entropy, but that is not captured by our method.

Our calculation is perturbative in the system-bath coupling strength $\kappa$. Note that the von Neumann entropy has a logarithmic singularity at the unperturbed state, which is pure. This is reflected by the fact that in addition to terms of order $\kappa^2$ (or any constant power of $\kappa$), terms of order $\kappa^{2s}$ (where $s$ is the number of replicas) play an important role.

\subsection{The model and general formulas}

Let us consider a quantum system (meant to represent a black hole) with some Hilbert space $\calH_\sys$ and Hamiltonian $H_\sys$. For an exact analogy with the evaporation problem, we would have to pick a pure state that looks like thermal to all simple measurements. Instead, we double the system and postulate that its initial state is the thermofield double, $\ket{\TFD_\sys}\in \calH_\sys^*\otimes\calH_\sys$. Only the right part is coupled to the heat bath, but we are interested in the von Neumann entropy of the double system as its density matrix $\rho_{\sys^*\sys}$ evolves in time. Likewise, the bath is also doubled, so that the initial state of the world is
\begin{equation}
\ket{\Psi_0} = \ket{\TFD_\sys} \otimes \ket{\TFD_\bath}
\in \calH_\sys^*\otimes\calH_\sys \otimes \calH_\bath\otimes\calH_\bath^*.
\end{equation}
A similar, but not identical,\footnote{In Refs.~\cite{GLQ17,CQZ20}, the initial state is taken to be the thermofield double of two \emph{interacting} subsystems rather than the product of two thermofield doubles.} setting was used in\cite{GLQ17,CQZ20}, where the $s$-Renyi entropy for integer $s>1$ was calculated.

The full Hamiltonian $H=H_\sys+H_\bath+H_{\sys\bath}$ acts only on the two objects in the middle, i.e.\ on $\calH_\sys\otimes\calH_\bath$. We assume that the interaction term has the form
\begin{equation}
H_{\sys\bath}  = \kappa \sum_{j=1}^{N} O_\sys^{j} O_\bath^j
\end{equation}
with some bosonic operators $O_\sys^{j}$ and $O_\bath^{j}$. In the case of fermionic systems like the SYK model~\cite{SaYe93,Kit.KITP.1,MS16,SoftMode}, we should multiply the coupling parameter $\kappa$ by $i$. For simplicity, we will do the computation for a bosonic system, but the final answer will equally be applicable to fermionic systems.

Thus, the evolution of the world in the interaction picture takes the form 
\begin{equation}
\begin{aligned}
\rho_{\sys^*\sys\bath\bath^*}(t)
=U(t)  \ket{\Psi_0}  \bra{\Psi_0} U^{-1}(t),\qquad U(t)
= \TT \left( e^{-i \int_0^t H_{\sys\bath} (u) du} \right),
\end{aligned}
\end{equation}
where $\TT$ stands for time ordering. We also assume that $\langle O_\sys^j \rangle = \langle O_\bath^j \rangle = 0$.\footnote{In case that $\langle O^j \rangle \neq 0$, one can work with $O^j - \langle O^j \rangle$.} In the rest of the section, we will compute the $s$-Renyi entropy of the system's density matrix $\rho_{\sys^*\sys}(t)$ after tracing out the bath. It is given by the perturbative expansion
\begin{equation} \label{ds}
\begin{aligned}
\rho_{\sys^*\sys}(t)= \sum_{n,m} \frac{(i\kappa)^n (-i\kappa)^m}{n!\, m!}
\int_0^t \biggl[
& \TT \{O^{j_n}_\sys(u_n) \cdots O^{j_1}_\sys(u_1)\} \bket{\TFD_\sys}
\bbra{\TFD_\sys} \tTT \{O^{j'_1}_\sys(u'_1) \cdots O^{j'_m}_\sys(u'_m)\} \\
&{} \times \Bigl\langle
\tTT \{O^{j'_1}_\bath(u'_1) \cdots O^{j'_m}_\bath(u'_m)\}
\TT \{O^{j_n}_\bath(u_n) \cdots O^{j_1}_\bath(u_1)\}
\Bigr\rangle
\biggr]\,du\,du',
\end{aligned}
\end{equation}
where the expectation value $\langle\cdots\rangle$ is with respect to the bath's thermal state and $\tTT$ denotes reverse time ordering. (If $u_1<\cdots<u_n$ and $u'_1<\cdots<u'_m$, then the operators are already ordered.) There is also an implicit sum over repeated indices, with each index going from $1$ to $N$. Note that operators with same indices have the same time argument.

\begin{figure}[t]
\centering
\begin{tabular}{c@{\hspace{2cm}}c}
\includegraphics{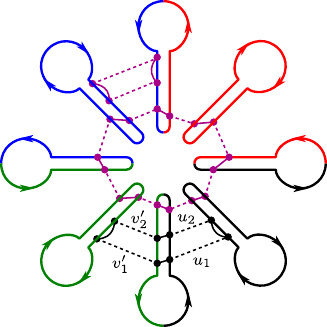} & \includegraphics{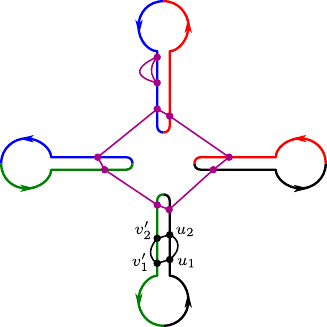}\\[3pt]
(a) & (b)
\end{tabular}
\caption{ (a) A diagram for $\Tr\left(\rho_{\sys^*\sys}(t)\right)^s$, where $\rho_{\sys^*\sys}(t)$ is given by \eqref{ds} and $s=4$. The thick solid lines make up $2s$ Keldysh contours for the system and the bath in alternating order. Each half of a contour represents a thermofield-double. Lorentzian time is directed toward the center of the diagram. Different replicas are depicted in different colors. Wick contractions of fields are depicted by thin solid lines, while the dotted lines represent system-bath coupling. The leading contributions are due to purple loops; the shorter and the longer loops correspond to equations \eqref{w0} and \eqref{w1}, respectively. (b) The simplified diagram obtained by omitting the bath's replicas.}
\label{fig:tcm}
\end{figure}

Since the combinatorics might soon get complicated, let us introduce some simplifying graphic notation:
\begin{equation}
\ket{\TFD_\sys}=\,\,\figbox{1.0}{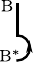}\,\,,\qquad\quad
\bra{\TFD_\sys}=\,\,\figbox{1.0}{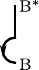}\,\,.
\end{equation}
Then each term in the expansion \eqref{ds} will look like this (where $\TT$, $\tTT$, and the indices are omitted):
\begin{equation}
O(u_2)O(u_1)\bket{\TFD_\sys}\,
\bcorr{O(u'_1)O(u_2)O(u_1)}_\bath
\bbra{\TFD_\sys} O(u'_1)=\,
\figbox{1.0}{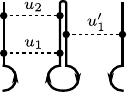}\qquad\quad
\text{for } u_1<u_2.
\end{equation}
The diagram element in the middle is the Keldysh contour for the heat bath. It consists of a circle at the bottom representing imaginary-time evolution and a stem corresponding to the real-time evolution; the time goes up. For integer $s$,\,  $\Tr\left(\rho_{\sys^*\sys}(t)\right)^s$ can be represented by gluing $s$ such diagrams (describing different replicas of the density matrix) in the cyclic order --- see  figure~\ref{fig:tcm}, where the replicas are depicted with different colors. The expectation values should be independently computed for each closed contour, whether it corresponds to the system or the bath.

Let us further assume that the system-bath coupling is sufficiently weak. Then the ``radiation quanta'' emanating from the system are sparse, which means that dominant diagrams have at most two operators with close times per contour. Therefore, the calculation can be done using Wick contraction. Of course, if the fields $O^j_\sys$, $O^j_\bath$ are Gaussian, then no sparseness condition is necessary.

An example of a (subleading) Wick pairing contributing to $\Tr\left( \rho_{\sys^*\sys}(t)\right)^s$ is as follows:
\begin{equation} \label{contract}
\wick{
\color{DarkGreen} \bigl\langle \c O(v'_1)
\color{black} \c {\color{DarkGreen}O}
\color{DarkGreen} (v'_2) \bigr\rangle_\bath
\bigl\langle \c2 O(v'_1) \c1 O(v'_2)
\color{black} \c1 O(u_2) \c2 O(u_1) \bigr\rangle_\sys
\bigl\langle \c O(u_2) \c O(u_1) \bigr\rangle_\bath
}\qquad\quad
\text{for }\, u_1<u_2,\quad v'_1<v'_2.
\end{equation}
It corresponds to the black loop at the bottom of figure~\ref{fig:tcm}\,(a). In general, a Wick contraction diagram is a disjoint union of loops that consist of alternating solid and dotted lines. Solid lines represent contractions of fields on the same contour, whereas dotted lines correspond to interaction terms such as $O^j_\sys(u) O^j_\bath(u)$. Each contraction comes with a Kronecker delta identifying the indices of the contracted fields. The result is nonzero if all the indices on each loop are the same. Therefore, each loop with $d$ solid line segments evaluates to $N \kappa^d$ multiplied by the product of two-point functions.

The diagrams for $\ln\bigl(\Tr \left(\rho_{\sys^*\sys}(t)\right)^s \bigr)$ are connected, i.e.\ contain a single loop. The factor of $N$ that appears here is usual for extensive thermodynamic quantities; thus, the intensive parameter is $\kappa^d$. If $\kappa$ is small, we should only keep loops with $d=2$ contractions (in one replica) and $d=2s$ contractions (traversing all the replicas). Both types of loops are shown in purple in figure~\ref{fig:tcm}.

There are three types of two-point functions for the system,
\begin{gather}
\label{GTdef}
\bcorr{\TT\{O^j_\sys(u)O^l_\sys(v)\}}=iG_{\TT\sys}(u-v)\delta_{jl},\qquad
\bcorr{\tTT\{O^j_\sys(u)O^l_\sys(v)\}}=iG_{\tTT\sys}(u-v)\delta_{jl},\\[3pt]
\label{Gdef}
\bcorr{O^{j'}_\sys(u')O^j_\sys(u)}=iG_{\sys}(u'-u)\delta_{j'j},
\end{gather}
and similarly for the bath. The expressions for closed paths will be simplified if we think of a two-point function as the matrix element of a bilocal operator $\widehat G$,
\begin{equation}
\bra{u'}  \widehat{G} \ket{u}=  f_t(u') G (u'-u) f_t(u),\qquad f_t(u) \equiv \theta(u) - \theta(u-t),
\end{equation}
where $f_t(u)$ is a time window function that vanishes outside the interval $(0,t)$. This way, we can extend the time domain to $(-\infty,\infty)$ and avoid putting limits on the integrals. We also define the transpose of the operator $\widehat{G}$, denoted by $\widehat{G}^{\trans}$, with the matrix element $\bra{t_1} \widehat{G}^{\trans} \ket{t_2} = G(t_2-t_1)$. Note that $\wh G_{\TT}$ and $\wh G_{\tTT}$ for bosons are symmetric, i.e.\ equal to their transpose. To illustrate this notation, the expression \eqref{contract} involves a loop with four Wick contractions, $v'_1 \xleftarrow{\bath} v'_2 \xleftarrow{\sys} u_2 \xleftarrow{\bath} u_1 \xleftarrow{\sys} v'_1$; hence, the result takes the form $\Tr \bigl(\widehat{G}_{\tTT\bath} \circ \widehat G_\sys \circ  \widehat{G}_{\TT\bath} \circ {\widehat{G}}^{\trans}_\sys \bigr)$.

As already mentioned, there are two types of loops that contribute to $\ln\bigl(\Tr \left(\rho_{\sys^*\sys}(t)\right)^s \bigr)$ to leading order. The loops of length $d=2$ can themselves be of two forms, one of which appears in figure~\ref{fig:tcm}. They give the following contributions:
\begin{equation} \label{w0}
P_{2,\TT}(t) = N\kappa^2 \,
\Tr \bigl(\wh G_{\TT\sys} \circ \wh{G}_{\TT\bath}^{\trans} \bigr),\qquad
P_{2,\tTT}(t) =  N\kappa^2 \,
\Tr \bigl(\wh G_{\tTT\sys}\circ \wh G_{\tTT\bath}^{\trans} \bigr).
\end{equation}
Here we have used the fact that $(\pm i)^2$ from \eqref{ds} cancels $i^2$ from \eqref{GTdef}. (The extra factors present in the fermionic case also cancel each other.) There are also loops of length $d=2s$, which we say to have winding number $1$ because they traverse all replicas. The expression for such a loop takes the following form:
\begin{equation} \label{w1}
P_{2s}(t) = N \kappa^{2s}\,
\Tr \bigl(-\wh G_\sys \circ \wh{G}_\bath^{\trans} \bigr)^s.
\end{equation}
Since in the end we are interested in the limit $s\rightarrow 1$, the quantities \eqref{w1} and \eqref{w0} are of the same order in the coupling.

As an exercise, let us sum up the leading diagrams in $\Tr\left(\rho_{\sys^*\sys}(t)\right)^s$ --- we should get the exponential of a sum of single loops with certain coefficients. It is sufficient to only keep track of the system's replica, leaving the bath implicit as in figure~\ref{fig:tcm}\,(b). Let $m_1,\dots,m_s$ and $n_1,\dots,n_s$ be the numbers of fields in the time ordered and anti-time ordered branches of the Keldysh contours. The number of ways to break them into $k$ loops of length $2s$ with winding number $1$ and some loops of length $2$ (with winding number $0$) is given by
\begin{equation}
\left(k! \right)^{2s-1} \prod_{r=1}^{s} \, \binom{m_r}{k}\binom{n_r}{k} \frac{(m_r-k)!}{\left(\frac{m_r-k}{2}\right)!\, 2^{\frac{m_r-k}{2}}}
\frac{(n_r-k)!}{\left(\frac{n_r-k}{2}\right)!\, 2^{\frac{n_r-k}{2}}}
\end{equation}
Defining $m_r - k =2 p_r$ and $n_r - k = 2q_r $, after manipulation we will get 
\begin{equation}
\begin{aligned} 
\Tr \left(\rho_{\sys^*\sys}(t)\right)^s 
&=\sum_{k=0}^{\infty} \sum_{p_1,\dots,p_s} \sum_{q_1,\dots,q_s}
\frac{1}{k!} \, \prod_{r=1}^{s} \, \frac{1}{(p_r)!(q_r)! \, 2^{p_r} 2^{q_r}} \,\,\bigl(P_{2s}(t) \bigr)^k \, \bigl(P_{2,\TT}(t) \bigr)^{p_r} \bigl(P_{2,\tTT}(t)\bigr)^{q_r} \\
 &= \exp \left(\frac{1}{2} \left(P_{2,\TT}(t)+P_{2,\tTT}(t)\right)+P_{2s}(t) \right)
\end{aligned}
\end{equation}
Using the fact that $\Tr\bigl(\wh{G}_{\TT\sys} \circ \wh{G}_{\TT\bath}^\trans\bigr) + \Tr\bigl(\wh{G}_{\tTT\sys} \circ \wh{G}_{\tTT\bath}^\trans\bigr) = 2 \Tr\bigl(\wh{G}_{\sys} \circ \wh{G}_{\bath}^\trans\bigr)$, the final answer is as follows:
\begin{equation} \label{mf}
\wideboxed{
\ln\Bigl(\Tr \left(\rho_{\sys^*\sys}(t)\right)^s \Bigr)
=N \biggl( s\kappa^2 \, \Tr \bigl( \widehat{G}_\sys \circ \widehat{G}_{\bath}^{\trans} \bigr)
+ \kappa^{2s} \Tr  \Bigl(\bigl(-\wh G_\sys \circ \wh{G}_\bath^{\trans}\bigr)^s \Bigr) \biggr).
}
\end{equation}
While the above equation was derived for bosonic systems, it is the same for fermionic systems like the SYK model.

\subsection{Short initial period vs.\ linear growth}

Equation \eqref{mf} allows one to compute the von Neumann entropy $S(\rho_{\sys^*\sys}(t))$ for $t<t_{\mathrm{Page}}$. Although the exact answer is model-dependent, there are two universal regimes: very early times, just after the system-bath coupling is turned on, and intermediate times, when the entropy grows linearly.\medskip

\noindent\textbf{Very early times:}\, Let $t \ll t_{\UV}$ such that the effect of the Hamiltonians $H_\sys$ and $H_\bath$ is negligible. For example, $t_{\UV}=J^{-1}$ for the SYK model. More exactly, we assume that the Green functions $G_{\sys}(t)$ and $G_{\bath}(t)$ may be approximated by some constants. Then the expression \eqref{mf} and the von Neumann entropy take this form:
\begin{gather}
\ln \Bigl(\Tr\left(\rho_{\sys^*\sys}(t)\right)^{s} \Bigr)
\approx N \Bigl( -s\, c\kappa^2t^2 +  \bigr(c\kappa^2t^2\bigr)^s \Bigr),
\\[3pt]
\wideboxed{
S\bigl(\rho_{\sys^*\sys}(t)\bigr) \approx c \kappa^2 t^2\,
\bigl(-\ln(c\kappa^2 t^2)+1 \bigr),\qquad
\text{where}\quad c=-G_{\sys}(0)G_{\bath}(0).
}
\medskip
\end{gather}

\noindent\textbf{Intermediate times:}\, For systems with continuous excitation spectrum, connected correlators decay in time. Exponential decay is typical; for example, in a conformal system at finite temperature, the correlator of fields with scaling dimension $\Delta$ decays as $\exp\bigl(-\frac{2\pi\Delta}{\beta}t\bigr)$ if $t\gg\frac{\beta}{\Delta}$. Let us assume that both $G_\sys(t)$ and $G_\bath(t)$ decay exponentially at $t\gg t_*$.

If $t\gg t_*$, then $\Tr\bigl(\wh{G}_\sys\circ \wh{G}_\bath^{\trans}\bigr)$ can be approximated as follows. This expression is as integral over $u,u'\in(0,t)$, but the integrand is negligible unless $|u'-u|\lesssim t_*$. Therefore, we may remove the limits on $u'$, and then use the Fourier transform:
\begin{equation}
\Tr\bigl(\wh{G}_\sys\circ \wh{G}_\bath^{\trans}\bigr)
\approx \int_{0}^{t} du \int_{-\infty}^{\infty}du'\,G_\sys(u',u)G_\bath(u',u)
=t\int \tG_\sys(\omega)\tG_\bath(-\omega)\,\frac{d\omega}{2\pi},
\end{equation}
where $\tG(\omega)=\int_{-\infty}^{\infty} G(t)e^{i\omega t}\,dt$. The second term in \eqref{mf} can also be approximated in such a way. Thus,
\begin{equation}
\ln \Bigr(\Tr(\rho(t)_{\sys^*\sys})^{s} \Bigr) \approx N A(s)\,t,
\end{equation}
where
\begin{equation}
A(s) = s\kappa^2 \int
\tG_\sys(\omega) \tG_\bath(-\omega)\, \frac{d\omega}{2\pi}
+ \kappa^{2s}  \int
\bigl(-\tG_\sys(\omega) \tG_\bath(-\omega)\bigl)^{s}\, \frac{d\omega}{2\pi}.
\end{equation} 
After analytically continuing to $s=1$, the von Neumann entropy will be given by
\begin{empheq}[box=\widebox]{gather}
S\bigl(\rho_{\sys^*\sys}(t)\bigr) \approx -NA'(1)\,t,\\[3pt]
\label{A1}
A'(1) = \kappa^2 \int \tG_\sys(\omega) \tG_\bath(-\omega)
\biggl(-\ln\Bigl(-\kappa^2\tG_\sys(\omega) \tG_\bath(-\omega)\Bigr)+1\biggr)
\frac{d\omega}{2\pi}.
\end{empheq} 

The integrals in $A(s)$ and $A'(1)$ converge because a possible peak at $\omega=0$ is broadened to have a width $\omega_{\mathrm{min}}=t_*^{-1}$. There is also a natural UV cutoff at $\omega_{\mathrm{max}} = t_{\UV}^{-1}$. An interesting case is where both the system and the bath are conformal at $\omega\ll\omega_{\mathrm{max}}$. Let us first assume that the temperature is zero; then $\tG_{\sys}(\omega)\propto\omega^{2\Delta_{\sys}-1}$ and $\tG_{\bath}(\omega)\propto\omega^{2\Delta_{\bath}-1}$ for $\omega>0$, but both $\tG_\sys$ and $\tG_\bath$ vanish at $\omega<0$. (Recall that these are Wightman functions.) Hence, $\tG_\sys(\omega) \tG_\bath(-\omega)$ is zero for all $\omega\not=0$. At finite temperature, the integral in \eqref{A1} is dominated by the region $\omega\sim\beta^{-1}$, where $i\tG_{\sys}(\omega)\sim t_{\UV} (t_{\UV}/\beta)^{2\Delta_{\sys}-1}$ and $i\tG_{\bath}(\omega)\sim t_{\UV} (t_{\UV}/\beta)^{2\Delta_{\bath}-1}$. It follows that
\begin{equation}
\frac{d S(\rho_{\sys^*\sys}(t))}{N\,dt}
=-A'(1)\sim \frac{-x\ln x}{\beta},\qquad
\text{where}\quad x=(\beta\kappa)^2
\biggl(\frac{\beta}{t_{\UV}}\biggr)^{-2(\Delta_\sys+\Delta_\bath)}.
\end{equation}

A good example is two SYK models at large $\beta J$. (A bath with $\Delta_\bath=\frac{1}{2}$ can also be realized by a critical Majorana chain.) The Renyi entropies in this case were studied in~\cite{CQZ20} using the effective action method, which is generally more powerful than perturbation theory. However, the analytic continuation to $s=1$ was not obtained. The computed growth rate of the $s$-Renyi entropy for integer $s>1$ is consistent with our estimate, 
\begin{equation}
\frac{d S_s(\rho_{\sys^*\sys}(t))}{N\,dt}
=\frac{A(s)}{1-s}\sim \beta\kappa^2
(\beta J)^{-2(\Delta_\sys+\Delta_\bath)}.
\end{equation}

\section{Perturbations to the saturated phase}\label{sec:motivation}

\begin{figure}
\centering \includegraphics{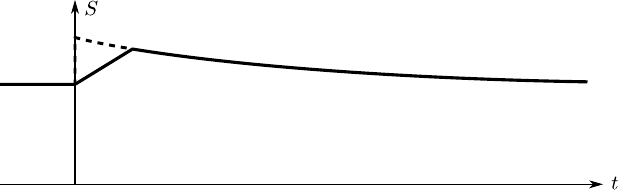}
\caption{Qualitative plot of the system's coarse-grained entropy (dashed line) and the entanglement entropy (solid line) in the presence of an instantaneous perturbation.}
\label{fig:entropy_plot}
\end{figure}

We now consider the system at later times, such that its von Neumann entropy has reached the coarse-grained (thermodynamic) entropy. The entanglement entropy in this phase shows interesting behavior under perturbations. For example, a short impulse increasing the system's energy (similar to throwing a rock into a black hole) will cause a resurgence of entropy growth. Indeed, such an action can be described by some unitary operator $V$. It increases the coarse-grained entropy and effective temperature, though the true microscopic entropy does not change. Letting the system interact with the bath, we should see a behavior similar to the cusp in the Page curve. Specifically, we expect the von Neumann entropy to grow until it becomes equal to the coarse-grained entropy. Since the growth rate is constant while the perturbation can be arbitrarily weak, the resurgence can be short --- just slightly longer than the scrambling time. It will be followed by a thermal equilibration period, when both the coarse-grained and microscopic entropies decrease, see figure~\ref{fig:entropy_plot}. Thus, the Page curve cusp is accessible in this setting. However, to actually produce a cusp, the perturbation should be sufficiently strong, likely beyond the Taylor expansion. We will study a simpler problem, calculating the effect in the lowest order.

While our formal goal is to compute the von Neumann entropy in a rather general setting, the key result pertains to systems that saturate the chaos bound~\cite{MSS15}. The expression obtained in this case admits a holographic interpretation, which will be discussed section~\ref{sec:discussion}.

\subsection{Statement of the problem}\label{sec:sf_problem}

For the study of the saturated phase, it is sufficient to consider one copy of the system and the bath rather than the thermofield double. The bath can be integrated out, giving rise to the interaction function $\sigma(\tau_1,\tau_2)=\kappa^2G_{\bath}(\tau_1,\tau_2)$ on the Keldysh contour. Let us simplify the model a bit and replace the interaction $\sigma$, which is constantly on, with a superoperator $R$ acting at a specific time. (The results we will obtain in this setting can be easily generalized to the original model.) So, the exact problem involves a quantum system at thermal equilibrium subjected to a sequence of two instantaneous perturbations, $V$ and $R$.

Let $V=e^{-ixX}$, where $X$ is a Hermitian operator and $x$ is a small parameter. We consider the action of $V$ on the thermal state $\rho_0$ and expand the resulting density matrix $\rho_1(x)$ to the second order in $x$:
\begin{equation}
\rho_1(x)=V\rho_0 V^{\dag}
\approx \rho_0 -ix[X,\rho_0]
+x^2\left(X\rho_0 X-\frac{1}{2}\bigl(X^2\rho_0+\rho_0X^2\bigr)\right).
\end{equation}
In fact, a nontrivial effect will be seen in the second order, and only when combined with a subsequent interaction with the environment. The latter is described by a physically realizable (i.e.\ completely positive, trace-preserving) superoperator $R$. Suppose that $R$ is close to the identity such that it can be expanded to the first order in some parameter $\epsilon$:
\begin{equation}\label{Lindblad}
R\approx 1\cdot 1+\epsilon L,\qquad
L =-i(C\cdot 1-1\cdot C)
+\sum_{j}\left(A_{j}\cdot A_{j}^{\dag}
-\frac{1}{2}\bigl(A_{j}^{\dag}A_{j}\cdot 1
+1\cdot A_{j}A_{j}^{\dag}\bigr)\right),
\end{equation}
where $A\cdot B$ stands for the superoperator that takes $\rho$ to $A\rho B$. The first term in $L$ (which involves a Hermitian operator $C$ and act as $\rho\mapsto -i[C,\rho]$) may be neglected because it represents an infinitesimal unitary transformation, and thus, does not change the entropy. The sum over $j$ (known as Lindbladian) corresponds to tracing out the environment. As will be justified later, we may replace $L$ with $\sum_{j}A_{j}\cdot A_{j}^{\dag}$ so that the final density matrix becomes
\begin{equation}\label{rho_xw}
\begin{aligned}
\rho(x,\epsilon)=R(\rho_1(x))
={}& (1-ix X)\rho_0(1+ix X) + \epsilon\sum_{j}
A_{j}(1-ix X)\rho_0(1+ix X)A_{j}^{\dag}\\
& +\text{unimportant terms}.
\end{aligned}
\end{equation}
Our goal is to compute $\frac{\partial^2}{\partial x^2} \frac{\partial}{\partial\epsilon} S(\rho(x,\epsilon))$, where $S(\rho)=-\Tr(\rho\ln\rho)$.\smallskip

We assume that $V$ acts at time $0$, whereas $R$ acts at a later time $t$. Thus, $A_j$ is understood as $A_j(t)=e^{iH_0t}A_j(0)e^{-iH_0t}$, where $A_j(0)$ is some simple (e.g.\ one- or two-body) operator. The calculation will be done by the replica method for a general large $N$ system in the early time regime, i.e.\ before the scrambling time. However $t$ is taken to be sufficiently large such that OTOCs are parametrically greater than correlators with non-alternating times. Note that $\rho(x,\epsilon)$ involves only non-alternating operators such as $A_{j}X\rho_0 XA_{j}^{\dag}$. However, OTOCs appear due to the use of replicas. The ``unimportant terms'' in \eqref{rho_xw} are exactly those that do not generate any OTOCs.

In the next section, we study partial derivatives of $S(\rho)$, assuming that $\rho$ depends on parameters in some particular way. This formalism has regular structure extending to higher orders, but it does not directly include the function $\rho(x,\epsilon)$ given by equation~\eqref{rho_xw}. To cover this case, we will use a trick called ``locking two operators in the same replica'', see section~\ref{sec:locking}. A streamlined entropy calculation, bypassing the general formalism, is given in section~\ref{sec:streamlined}.

\subsection{Thermodynamic response theory for the replicated system}\label{sec:definitions}

Let us recall the standard definition of connected correlators. We begin with the partition function $Z=\Tr W$, where $W$ is the imaginary-time evolution operator:
\begin{equation}\label{RZ}
W=\TT\exp\left(-\int_{0}^{\beta}H(\tau)\,d\tau\right).
\end{equation}
Without perturbation, we have $H(\tau)=H_0$. The insertion of operators $X_1,\dots,X_n$ at times $\tau_1,\dots,\tau_n$ is described by perturbing the Hamiltonian:
\begin{equation}
H(\tau)=H_0-\sum_{j=1}^{n}x_j\delta(\tau-\tau_j)\,X_j,\qquad\quad \beta\ge\tau_n\ge\dots\ge\tau_1\ge 0,
\end{equation}
where $x_j$ are infinitesimal numbers. We generally assume that the operators $X_j$ are bosonic. (If any of them is fermionic, the corresponding variable $x_j$ should be anti-commuting.) Thus,
\begin{equation}
W(\beta,x_n,\dots x_1)
= e^{-(\beta-\tau_n) H_0}(1+x_nX_n)e^{-(\tau_n-\tau_{n-1}) H_0}
\cdots (1+x_1X_1)e^{-\tau_1 H_0}
\end{equation}
and $Z(\beta,x_n,\dots x_1)=\Tr W(\beta,x_n,\dots x_1)$. The full correlator is simply
\begin{equation}
\corr{X_n(\tau_n)\cdots X_1(\tau_1)}
=\left.Z^{-1}\frac{\partial^n Z}{\partial x_1\cdots\partial x_n}\right|
_{x_1=\dots=x_n=0}.
\end{equation}
The corresponding connected correlator is defined as follows:\footnote{We use commas instead of double brackets because the usual notation (without commas) has some ambiguity.}
\begin{equation}
\corr{X_n(\tau_n),\dots ,X_1(\tau_1)}
=\left.\frac{\partial^n \ln Z}{\partial x_1\cdots\partial x_n}\right|
_{x_1=\dots=x_n=0}.
\end{equation}
For example, $\corr{X,Y}=\corr{XY}-\corr{X}\corr{Y}$ and
\begin{equation}\label{ccorr3}
\corr{X,Y,Z}=\corr{XYZ}
-\corr{XY}\corr{Z}-\corr{XZ}\corr{Y}-\corr{YZ}\corr{X}
+2\corr{X}\corr{Y}\corr{Z}.
\end{equation}

Now, let us introduce $s$ replicas of the system, such that the partition function becomes
\begin{equation}
Z(s,\beta,x_n,\dots,x_1) = \Tr\bigl(W(\beta,x_n,\dots,x_1)\bigr)^s.
\end{equation}
We may think of the parameter $s$ as being associated with a \emph{branching operator} $\BB$, which commutes with everything. It is not defined by itself but only through its connected correlators:
\begin{equation}\label{brcorr}
\wideboxed{
\corr{\BB,X_n(\tau_n),\cdots,X_1(\tau_1)}
= \left.\frac{\partial^n}{\partial x_1\cdots\partial x_n}\right|
_{x_1=\dots=x_n=0}
\left.\frac{\partial\ln Z}{\partial s}\right|_{s=1}.
}
\end{equation}
The branched correlator~\eqref{brcorr} is related to the entropy $S=S(\rho)$ of the density matrix $\rho=Z^{-1}W$ at $s=1$ because
\begin{equation}
\left.\bigl(\partial_s(\ln Z)\bigr)\right|_{s=1}=\ln Z-S.
\end{equation}
Thus, the entropy derivative with respect to $x_1,\dots,x_n$ is given by $-\corr{\BB,X_n(\tau_n),\cdots,X_1(\tau_1)} +\corr{X_n(\tau_n),\cdots,X_1(\tau_1)}$. It is usually the easiest to compute the derivative of the relative entropy, $S(\rho||\rho_0) =\Tr(\rho(\ln\rho-\ln\rho_0))$:
\begin{equation}
\begin{aligned}
\left.\frac{\partial^n S(\rho||\rho_0)}{\partial x_1\cdots\partial x_n}\right|
_{x_1=\dots=x_n=0}
&= \bcorr{(\BB+\beta H_0),X_n(\tau_n),\cdots,X_1(\tau_1)}
- \bcorr{X_n(\tau_n),\cdots,X_1(\tau_1)}\\
&=\left.\frac{\partial^n}{\partial x_1\cdots\partial x_n}\right|
_{x_1=\dots=x_n=0}
\left.\Bigl(\partial_{s}
\bigl(s^{-1}\ln\Tr\,(W(\beta/s,\ldots))^s\bigr)\Bigr)\right|_{s=1}.
\end{aligned}
\end{equation}
For integer $s$, the expression $\Tr\,(W(\beta/s,\ldots))^s$ may be interpreted in terms of gluing $s$ intervals of length $\beta/s$ to make a circle of length $\beta$. The operators $X_n(\tau_n),\dots,X_1(\tau_1)$ are distributed along that circle. Thus, the number in question is, essentially, a correlation function at the given $\beta$.

An important caveat is that there is no natural definition of the full correlator $\corr{\BB A}$ as a function of $A$ such that one could compute $\corr{\BB YZ}$ by substituting $YZ$ for $A$. If such a function (with the usual relation to the connected correlator) existed, we would have this corollary of equation~\eqref{ccorr3}: $\corr{\BB,Y,Z} =\corr{\BB,YZ}-\corr{\BB,Y}\corr{Z}-\corr{Y}\corr{\BB,Z}$. But this last identity is false because in the expression for $\corr{\BB,Y,Z}$, the operators $Y$ and $Z$ can occur in different replicas, but in $\corr{\BB,YZ}$, they cannot.

\subsection{Branched two-point correlator}\label{sec:two-point}

Suppose the ordinary correlation function $\corr{Y(\tau),X(0)}$ is known on the imaginary axis, $\tau=it$, and let us use its Fourier transform in $t$. In these terms,
\begin{equation}
\corr{Y(\tau),X(0)}
=\int_{-\infty}^{\infty} F_{Y,X}(\omega)e^{-\omega\tau}\,
\frac{d\omega}{2\pi}.
\end{equation}
The corresponding branched correlator is expected to have a similar form,
\begin{equation}
\bcorr{\BB+\beta H_0,Y(\tau),X(0)}-\bcorr{Y(\tau),X(0)}
=\int_{-\infty}^{\infty} h_{Y,X}(\omega)e^{-\omega\tau}\,
\frac{d\omega}{2\pi}.
\end{equation}
The goal of this section is to find the function $h_{Y,X}$.

Let us consider the Fourier modes of the operators $Y$ and $X$, for example, $Y_{\omega} =\int Y(it)e^{i\omega t}\,dt$. Their connected correlator is
\begin{equation}
\corr{Y_{\omega},X_{\omega'}}=F_{Y,X}(\omega)\cdot 2\pi\delta(\omega+\omega'),
\end{equation}
and we also have
\begin{equation}
Y(\tau)=\int\underbrace{Y_{\omega}e^{-\omega\tau}}_{Y_{\omega}(\tau)}\,
\frac{d\omega}{2\pi},\qquad
X(0)=\int\underbrace{X_{\omega}}_{X_{\omega}(0)}\,\frac{d\omega}{2\pi}.
\end{equation}
We now calculate the branched correlator of $Y_{\omega}$ and $X_{\omega'}$, equal to $h_{Y,X}(\omega)\cdot 2\pi\delta(\omega+\omega')$. When the number of replicas $s$ is a positive integer, each of the operators in question can be inserted in any replica, so the calculation involves a double sum. Since each replica's length is $\beta/s$, putting $Y_{\omega}$ in the $k$-th replica is described by $Y_{\omega}(k\beta/s)=Y_{\omega}e^{-k\beta\omega/s}$. With this in mind, we get:
\begin{align}
\hspace{30pt}&\hspace{-30pt}
\bcorr{\BB+\beta H_0,Y_{\omega},X_{\omega'}}
-\bcorr{Y_{\omega},X_{\omega'}}\\
\label{sum_kl}
&=\left.\left(\partial_s\left(s^{-1}\sum_{k=0}^{s-1}\sum_{l=0}^{s-1}\TT
\biggl\langle Y_{\omega}\biggl(\frac{k\beta}{s}\biggr),\,
X_{-\omega'}\biggl(\frac{l\beta}{s}\biggr)\biggr\rangle
\right)\right)\right|_{s=1}\\
&=\left.\left(\partial_s\sum_{k=0}^{s-1}
\biggl\langle Y_{\omega}\biggl(\frac{k\beta}{s}\biggr),\,
X_{-\omega'}(0)\biggr\rangle \right)\right|_{s=1}\\
&=\bcorr{Y_{\omega},X_{\omega'}}
\biggl(\partial_s
\ubrace{\sum_{k=0}^{s-1}e^{-k\beta\omega/s}}
_{\frac{1-u}{1-u^{1/s}}\:\,\text{for }u=e^{-\beta\omega}}
\biggr)\biggr|_{s=1}
=2\pi\delta(\omega+\omega')\cdot F_{Y,X}(\omega)
\frac{\beta\omega}{e^{\beta\omega}-1}.
\end{align}
Thus,
\begin{equation}
\wideboxed{
h_{Y,X}(\omega)=F_{Y,X}(\omega)\frac{\beta\omega}{e^{\beta\omega}-1}.
}
\end{equation}

\subsection{Branched correlator related to early-time OTOCs}

Let us recall the original problem of computing $S(\rho(x,\epsilon))$ with $\rho(x,\epsilon)$ given by equation~\eqref{rho_xw}. In this section, we calculate an analogous branched correlator $\corr{\BB, A_j(\beta+it),X(\beta), X(0),A_k^{\dag}(it)}$ and, more generally,
\begin{equation}
\bcorr{\BB,X_4(\beta+it_4),X_3(\beta+it_3),X_2(it_2),X_1(it_1)}\qquad
\text{for }\, t_1,t_4\approx t,\quad t_2,t_3\approx 0.
\end{equation}
One can eliminate $\beta$ from the time arguments by cyclically permuting $X_4,\ldots,X_1$. As already mentioned, the replica calculation involves OTOCs, which are dominant for sufficiently large $t$. Neglecting all terms with non-alternating times, we get:
\begin{equation}
\begin{aligned}
\Corr\bydef
\bcorr{\BB,X_4(\beta+it_4),X_3(\beta+it_3),X_2(it_2),X_1(it_1)}
&=\bcorr{\BB,X_2(it_2),X_1(it_1),X_4(it_4),X_3(it_3)}\\
&\approx \left.\Bigl(\partial_s\bigl(\Corr_{+}(s)+\Corr_{-}(s)\bigr)
\Bigr)\right|_{s=1},
\end{aligned}
\end{equation}
where
\begin{align}
\label{OTOC1}
\Corr_{+}(s) &=
\sum_{k=0}^{s-1} \sum_{j=0}^{k} \sum_{l=k+1}^{s-1} \biggl\langle
X_1\left(it_1+\frac{l\beta}{s}\right),
X_2\left(it_2+\frac{k\beta}{s}\right),
X_4\left(it_4+\frac{j\beta}{s}\right),
X_3(it_3)\biggr\rangle,
\\
\Corr_{-}(s) &=
\sum_{k=0}^{s-1} \sum_{j=0}^{k} \sum_{l=k+1}^{s-1} \biggl\langle
X_2(it_2),
X_1\left(it_1-\frac{j\beta}{s}\right),
X_3\left(it_3-\frac{k\beta}{s}\right),
X_4\left(it_4-\frac{l\beta}{s}\right)\biggr\rangle.
\end{align}
\begin{figure}[t]
\centering \includegraphics{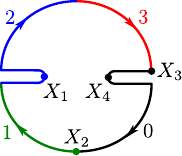}
\caption{Graphic representation of a single term in \eqref{OTOC1}. In this example, $s=4$ (with the replicas labeled $0,1,2,3$),\, $k=1$,\, $j=0$,\, $l=2$, and $t_2=t_3=0$.}
\label{fig:OTOC1}
\end{figure}
(Equation \eqref{OTOC1} is illustrated by figure~\ref{fig:OTOC1}.) In order to make further progress, we will use the single-mode ansatz for early-time OTOCs~\cite{SoftMode},
\begin{equation}
\bcorr{X_1(\tau_1),X_2(\tau_2),X_4(\tau_4),X_3(\tau_3)}
\approx -C^{-1}e^{-i\kap(\tau_1+\tau_4-\tau_2-\tau_3-\beta/2)/2}\,
\Upsilon^{\R}_{X_1,X_4}(\tau_1-\tau_4)
\Upsilon^{\A}_{X_2,X_3}(\tau_2-\tau_3),
\end{equation}
combined with the Fourier representation
\begin{equation}
\Upsilon^{\R}_{X_1,X_4}(\tau) =\int
\tilde{\Upsilon}^{\R}_{X_1,X_4}(\omega)e^{-\omega\tau}\,\frac{d\omega}{2\pi},
\qquad
\Upsilon^{\A}_{X_2,X_3}(\tau) =\int
\tilde{\Upsilon}^{\A}_{X_2,X_3}(\omega)e^{-\omega\tau}\,\frac{d\omega}{2\pi}.
\end{equation}
The result has this general form:
\begin{empheq}[box=\widebox]{align}
\label{BB1}
&\Corr=-C^{-1}e^{\kap(t_1+t_4-t_2-t_3)/2}\int
\tilde{\Corr}(\omega_{14},\omega_{23})\,
e^{-i\omega_{14}(t_1-t_4)}e^{-i\omega_{23}(t_2-t_3)}\,
\frac{d\omega_{23}}{2\pi}\frac{d\omega_{14}}{2\pi},\\[3pt]
\label{BB2}
&\tilde{\Corr}(\omega_{14},\omega_{23})=
\tilde{\Upsilon}^{\R}_{X_1,X_4}(\omega_{14})
\tilde{\Upsilon}^{\A}_{X_2,X_3}(\omega_{23})\,
f(\omega_{14},\omega_{23}),
\end{empheq}
so the task is to compute $f(\omega_{14},\omega_{23})$.

First, we find the similar function $f_{+}(s;\omega_{14},\omega_{23})$ related to $\Corr_{+}(s)$. Let
\begin{equation}\label{uvw}
u=e^{-\beta\omega_{23}},\qquad v=e^{-\beta\omega_{14}},\qquad
w=e^{-i\beta\kap/2}.
\end{equation}
Then
\begin{align}
&f_{+}(s;\omega_{14},\omega_{23})
=\sum_{k=0}^{s-1} \sum_{j=0}^{k} \sum_{l=k+1}^{s-1}
u^{k/s}v^{(l-j)/s}w^{(j+l-k)/s-1/2}\\
&\hspace{20pt}\begin{aligned}
=\frac{w^{-1/2}}{(1-(w/v)^{1/s})(1-(wv)^{1/s})}
\biggl(&(vw)^{1/s}\frac{1-u/v}{1-(u/v)^{1/s}}
-vw\frac{1-u/w}{1-(u/w)^{1/s}}\\
&-w^{2/s}\frac{1-uw}{1-(uw)^{1/s}}
+v^{1-1/s}w^{1+1/s}\frac{1-u/v}{1-(u/v)^{1/s}}\biggr),
\end{aligned}
\end{align}
and hence,
\begin{equation}
\begin{aligned}
\left.\bigl(\partial_{s}f_{+}(s;\omega_{14},\omega_{23})\bigr)\right|_{s=1}
={}&-\frac{1}{(1-uw)(1-uw^{-1})}
\left(\frac{w^{-1/2}}{1-v^{-1}u^{-1}}+\frac{w^{1/2}}{v^{-1}-u^{-1}}\right)
\ln u\\
&+\frac{1}{(1-v^{-1}w)(1-v^{-1}w^{-1})}
\left(\frac{w^{-1/2}}{1-uv}+\frac{w^{1/2}}{u-v}\right)
\ln v\\
&+\frac{w^{-1/2}(1+uv^{-1})-w^{1/2}(u+v^{-1})}
{(1-uw)(1-uw^{-1})(1-v^{-1}w)(1-v^{-1}w^{-1})}\,\ln w.
\end{aligned}
\end{equation}
The function $f_{-}$ is obtained from $f_{+}$ by replacing $w$ with $w^{-1}$. Adding both terms together, we get:
\begin{equation}\label{f1423}
\begin{aligned}
f(\omega_{14},\omega_{23})
&=
\frac{(w^{-1/2}+w^{1/2})(u^{-1}-1)(1+v^{-1})}
{(1-uw)(1-uw^{-1})(1-u^{-1}v^{-1})(v^{-1}-u^{-1})}\,\ln u\\
&+\frac{(w^{-1/2}+w^{1/2})(1+u)(1-v)}
{(1-v^{-1}w)(1-v^{-1}w^{-1})(1-uv)(u-v)}\,\ln v\\
&+\frac{(w^{-1/2}-w^{1/2})(1+u)(1+v^{-1})}
{(1-uw)(1-uw^{-1})(1-v^{-1}w)(1-v^{-1}w^{-1})}\,\ln w,
\end{aligned}
\end{equation}
where $u=e^{-\beta\omega_{23}}$,\, $v=e^{-\beta\omega_{14}}$, and
$w=e^{-i\beta\kap/2}$.

A great simplification occurs in the maximal chaos case:
\begin{equation}\label{f1423_simple}
\wideboxed{
f(\omega_{14},\omega_{23})
=\frac{2\pi}{(1+e^{-\beta\omega_{23}})(1+e^{\beta\omega_{14}})}\qquad
\text{if}\quad \kap=\frac{2\pi}{\beta}.
}
\end{equation}
Importantly, the function $f(\omega_{14},\omega_{23})$ splits into two factors. They may be interpreted in terms of interaction of the fluctuating horizon (which corresponds to $\BB$) with incoming and outgoing radiation, see section~\ref{sec:discussion}.

\subsection{Locking two operators in the same replica}\label{sec:locking}

We now adapt the obtained result to express the entropy of the density matrix $\rho(x,\epsilon)$. The latter is a normalized version of the operator
\begin{equation}\label{Wbar}
\overline{W}(\beta,x,\epsilon)
=(1-ixX)e^{-\beta H_0}(1+ixX) +\epsilon\sum_{j}
A_{j}(t)\,(1-ixX)e^{-\beta H_0}(1+ixX)\,A_{j}^{\dag}(t).
\end{equation}
Note that we have made the time explicit and will follow the convention that $A_j=A_j(0)$. The corresponding entropy could be calculated as in the previous section while restricting $X_1=A_j^\dag$ and $X_4=A_j$ to be in the same replica. However, instead of imposing this restriction by hand, we will modify the problem so that it fits the branched correlator setting. First, we replace the set of operators $A_j$ with a single operator $Y$. This is achieved by extending the physical system with an auxiliary one, comprising a ground state $\ket{0}$ with zero energy and a set of excited states $\ket{j}$ with energy $\Omega$. We denote the Hamiltonian of the extended system by $H(\Omega)$ and set
\begin{equation}
Y=\sum_{j}A_{j}\otimes\ket{0}\bra{j},\qquad
Y^{\dag}=\sum_{j}A_{j}^{\dag}\otimes\ket{j}\bra{0}.
\end{equation}
Although the transformation just described alters the operator $\overline{W}(\beta,x,\epsilon)$ in a nontrivial way, we will find an agreement in the $\Omega\to\infty$ limit. For the time being, let us construct some operators acting on the extended system that correspond to the two terms in~\eqref{Wbar} as closely as possible:
\begin{align}
&(1-ixX)e^{-\beta H(\Omega)}(1+ixX)
=(1-ixX)e^{-\beta H_0}(1+ixX)
\otimes\left(\ket{0}\bra{0}+e^{-\beta\Omega}\sum_{j}\ket{j}\bra{j}\right),
\\[2pt]
\label{YAj}
&\begin{aligned}
e^{\beta\Omega}\,Y(t)\,(1-ixX)e^{-\beta H(\Omega)}&(1+ixX)\,Y^{\dag}(t)\\
&=\left(\sum_{j}A_{j}(t)\,(1-ixX)e^{-\beta H_0}(1+ixX)\,A_{j}^{\dag}(t)\right)
\otimes\ket{0}\bra{0}.
\end{aligned}
\end{align}

Now let
\begin{equation}
W(\Omega,\beta,x,y)
=(1+e^{\beta\Omega/2}yY(t))(1-ixX)e^{-\beta H(\Omega)}(1+ixX)
(1+e^{\beta\Omega/2}yY^\dag(t)).
\end{equation}
Then
\begin{equation}
\lim_{\Omega\to\infty}W(\Omega,\beta,x,y)
=\overline{W}(\beta,x,y^2)\otimes\ket{0}\bra{0},
\end{equation}
and hence,
\begin{equation}\label{tildeWs}
\lim_{\Omega\to\infty}\Tr\bigl(W(\Omega,\beta,x,y)\bigr)^{s}
=\Tr\bigl(\overline{W}(\beta,x,y^2)\bigr)^s
\end{equation}
for any $s$. The last equation can be interpreted as the operators $Y(t)$ and $Y^\dag(t)$ in the expansion of $\Tr W^s$ being locked in the same replica.

We now take the $s$ derivative of both sides at $s=1$ and consider the $x^2y^2$ term in the Taylor expansion. Thus, the right-hand side of equation \eqref{tildeWs} becomes $\frac{1}{2} \partial_x^2 \partial_\epsilon\,S(\rho(x,\epsilon)) \big|_{x=\epsilon=0}$. As for the left-hand side, one should keep in mind that the two $\partial_x$ derivatives may act on $(1\pm ix X)$ factors with different signs in any pair of replicas or with the same sign in different replicas. These contributions give rise, respectively, to the first and the last two long lines in the equation below. The same-replica terms are subtracted anyway since we are computing derivatives of $S=\ln Z -\left.\partial_s(\ln Z)\right|_{s=1}$ rather than just $\left.\partial_s(\ln Z)\right|_{s=1}$. Thus, the result reads:
\begin{align}
\label{x2epsilonnew}
&\frac{1}{2}\partial_x^2
\partial_\epsilon\,S(\rho(x,\epsilon))
\bigr|_{x=\epsilon=0}\\
\nonumber
&\hspace{10pt}=\begin{aligned}[t]
\lim_{\Omega\to\infty}\biggl(&e^{\beta\Omega}\Bigl(
-\bcorr{\BB,Y(\beta+it),X(\beta),X(0),Y^{\dag}(it)}_{\Omega}
+\bcorr{Y(\beta+it),X(\beta),X(0),Y^{\dag}(it)}_{\Omega}\\
&+\frac{1}{2}\bcorr{\BB,Y(\beta+it),X(0),X(0),Y^{\dag}(it)}_{\Omega}
-\frac{1}{2}\bcorr{Y(\beta+it),X(0),X(0),Y^{\dag}(it)}_{\Omega}\\
&+\frac{1}{2}\bcorr{\BB,Y(\beta+it),X(\beta),X(\beta),Y^{\dag}(it)}_{\Omega}
-\frac{1}{2}\bcorr{Y(\beta+it),X(\beta),X(\beta),Y^{\dag}(it)}_{\Omega}
\Bigr)\biggr).
\end{aligned}
\end{align}
Neglecting all terms with non-alternating times and keeping those involving OTOCs, we are left with three branched correlators. We will compute them using the method of the previous section and observe massive cancellation. For a more illuminating calculation, see section~\ref{sec:streamlined}.

Let us consider the first correlator in more detail and establish a correspondence with our previous notation. Part of it is obvious: $X_1=Y^\dag$,\, $X_2=X_3=X$, and $X_4=Y$. In the OTOC-based approximation, the branched correlator
\begin{equation}
\Corr=\bcorr{\BB,Y(\beta+it),X(\beta),X(0),Y^{\dag}(it)}_{\Omega}
=\bcorr{\BB,X(0),Y^{\dag}(it),Y(it),X(0)}_{\Omega}
\end{equation}
is given by \eqref{BB1}, \eqref{BB2} with $t_1=t_4=t$ and $t_2=t_3=0$. Note that if $\Omega$ is large, then
\begin{equation}
\Upsilon^{\R}_{Y^\dag,Y}(\tau)
\approx e^{-\Omega(\beta-\tau)}
\sum_{j} \Upsilon^{\R}_{A_j^\dag,A_j}(\tau)
\end{equation}
by analogy with the rather obvious equation $\corr{Y^\dag(\tau)Y(0)} \approx e^{-\Omega(\beta-\tau)} \sum_{j}\corr{A_j^\dag(\tau)A_j(0)}$. Hence,
\begin{equation}\label{Upsilon_shift}
\tilde{\Upsilon}^{\R}_{Y^\dag,Y}(\omega)
\approx e^{-\beta\Omega}
\sum_{j}\tilde{\Upsilon}^{\R}_{A_j^\dag,A_j}(\omega+\Omega).
\end{equation}
On the other hand, the function $\tilde{\Corr}(\omega_{14},\omega_{23})$ in equation \eqref{BB1} may be replaced with $\tilde{\Corr}(\omega_{14}-\Omega,\omega_{23})$ without affecting the result.\footnote{This is because $t_1=t_4$. Note, however, that the same condition was implicitly used in \eqref{YAj}. A more general model of system-bath coupling involves $A_j(t_4)$ and $A_j^\dag(t_1)$ so that the additional factor $e^{-i\Omega(t_1-t_4)}$ has to be added on the left-hand side of \eqref{YAj}. To reproduce this factor, one \emph{should} replace $\tilde{\Corr}(\omega_{14},\omega_{23})$ with $\tilde{\Corr}(\omega_{14}-\Omega,\omega_{23})$ in \eqref{BB1}.} Combining \eqref{BB2} with \eqref{Upsilon_shift}, we get
\begin{equation}
e^{\beta\Omega}\tilde{\Corr}(\omega_{14}-\Omega,\omega_{23})
=\sum_{j}\tilde{\Upsilon}^{\R}_{A_j^\dag,A_j}(\omega_{14})
\tilde{\Upsilon}^{\A}_{X,X}(\omega_{23})\,
f(\omega_{14}-\Omega,\,\omega_{23}),
\end{equation}
where the prefactor $e^{\beta\Omega}$ is used to match the right-hand side of \eqref{x2epsilonnew}.

Thus, we have arrived at the conclusion that the replica locking amounts to replacing the function $f(\omega_{14},\omega_{23})$ in \eqref{BB2} with $f(-\infty,\omega_{23})$. Using the explicit formulas \eqref{f1423}, \eqref{f1423_simple}, we get:
\begin{equation}\label{f_inf}
\begin{aligned}
f(-\infty,\omega_{23}) &=\frac{
\cos\frac{\beta\kap}{4}\cdot 2\beta\omega_{23}(1-e^{-\beta\omega_{23}})
+\sin\frac{\beta\kap}{4}\cdot \beta\kap(1+e^{-\beta\omega_{23}})}
{(1-e^{-\beta(\omega_{23}+i\kap/2)})(1-e^{-\beta(\omega_{23}-i\kap/2)})}\\[3pt]
&=\frac{2\pi}{1+e^{-\beta\omega_{23}}}\qquad
\text{if}\quad \kap=\frac{2\pi}{\beta}.
\end{aligned}
\end{equation}

Now, we consider the other two terms in~\eqref{x2epsilonnew}. These contributions are recognized as special cases of
\begin{equation}
\begin{aligned}
\mathcal{B}_2:&=\frac{1}{2}\bcorr{\BB,X_4(\beta+it_4),X_2(it_2),X_2(it_2),X_1(it_1)}\approx \partial_s\mathcal{B}_2(s)|_{s=1},\\
\mathcal{B}_3:&=\frac{1}{2}\bcorr{\BB,X_4(\beta+it_4),X_3(\beta+it_3),X_3(\beta+it_3),X_1(it_1)}\approx \partial_s\mathcal{B}_3(s)|_{s=1},
\end{aligned}
\end{equation}
where we have introduced
\begin{equation}
\begin{aligned}
\mathcal{B}_2(s)&=\sum_{k=1}^{s-1}\sum_{j=0}^{k-1}\sum_{l=k}^{s-1}\left< X_2\left(it_2\right)X_1\left(it_1-\frac{j\beta}{s}\right)X_2\left(it_2-\frac{k\beta}{s}\right)X_4\left(it_4-\frac{l\beta}{s}\right)\right>,\\
\mathcal{B}_3(s)&=\sum_{k=1}^{s-1}\sum_{j=0}^{k-1}\sum_{l=k}^{s-1}\left< X_1\left(it_1+\frac{l\beta}{s}\right)X_3\left(it_3+\frac{k\beta}{s}\right)X_4\left(it_4+\frac{j\beta}{s}\right)X_3(it_3)\right>.
\end{aligned}
\end{equation}
We proceed with the calculation and apply the single-mode ansatz for OTOCs. It gives
\begin{equation}
\mathcal{B}_j=-C^{-1}e^{\varkappa(t_1+t_4-2t_j)/{2}}\int \tilde\Upsilon^\text{R}_{X_1X_4}(\omega_{14})\tilde\Upsilon^\text{A}_{X_jX_j}(\omega_{23})f_j(\omega_{14},\omega_{23})e^{-i\omega_{14}(t_1-t_4)}\frac{d\omega_{23}}{2\pi}\frac{d\omega_{14}}{2\pi},
\end{equation}
with $f_j(\omega_{14},\omega_{23})=\partial_sf_j(s,\omega_{14},\omega_{23})|_{s=1}$. For $j=2$, we find that
\begin{equation}
f_2(s,\omega_{14},\omega_{23})=\sum_{k=1}^{s-1}\sum_{j=0}^{k-1}\sum_{l=k}^{s-1}u^{k/s}v^{(l-j)/s}w^{-(j+l-k)/s+1/2},
\end{equation}
where the variables $(w,v,w)$ are defined as in~\eqref{uvw}. This leads to the expression
\begin{equation}
\begin{aligned}
f_2(\omega_{14},\omega_{23})=&\frac{u^2 v w^{3/2} (-u (v+w)+v w+1)}{(u-v) (u v-1) (u-w) (u w-1)}\ln u+\frac{u v^2 w^{3/2} (u (v-w)+v w-1)}{(u-v) (u v-1) (v-w) (v w-1)} \ln v\\
&+\frac{u v w^{5/2} (-w (u+v)+u v+1)}{(u-w) (u w-1) (v-w) (v w-1)}\ln w.
\end{aligned}
\end{equation}
The function $f_3$ is obtained from $f_2$ by replacing $w$ with $w^{-1}$. Adding these new contributions to $f(\omega_{14},\omega_{23})$ in~\eqref{f1423} or~\eqref{f_inf}, we arrive at the following result: 
\begin{equation}
\frac{1}{2}\left.{\partial_x^2}
\partial_\epsilon\,S(\rho(x,\epsilon))
\right|_{x=\epsilon=0}=C^{-1}e^{\varkappa t}\int \sum_j\tilde\Upsilon^\text{R}_{A^\dagger_jA_j}(\omega_{14})\tilde\Upsilon^\text{A}_{XX}(\omega_{23})f_S(\omega_{14},\omega_{23})\frac{d\omega_{23}}{2\pi}\frac{d\omega_{14}}{2\pi},
\end{equation}
where 
\begin{equation}\label{result}
\begin{aligned}
f_S(\omega_{14},\omega_{23})&=f(-\infty,\omega_{23})-f_2(-\infty,\omega_{23})-f_3(-\infty,\omega_{23})\\
&=2\beta \omega_{23}\cos\left(\frac{\varkappa \beta}{4}\right)+\beta \varkappa\sin\left(\frac{\varkappa \beta}{4}\right)\\
&=2\pi\qquad
\text{if}\quad \varkappa=\frac{2\pi}{\beta}.
\end{aligned}
\end{equation}

\subsection{Direct calculation of the entropy}\label{sec:streamlined}

The simplicity of equation~\eqref{result} indicates cancellation between different branched four-point correlators. The remaining terms contain pairs of operators in the same or adjacent replicas. To understand this better, let us recall some general points explained earlier: the change in entanglement entropy under the perturbation is proportional to $\frac{1}{2} \partial_x^2 \partial_\epsilon\,S(\rho(x,\epsilon)) \big|_{x=\epsilon=0}$, where $S=\ln Z -\left.\partial_s(\ln Z)\right|_{s=1}$ with $Z=\Tr\overline{W}^s$ and
\begin{equation}\label{Wbar1}
\overline{W}(\beta,x,\epsilon)
=e^{-ixX}e^{-\beta H_0}e^{ixX} +\epsilon\sum_{j}
A_{j}(it)\,e^{-ixX}e^{-\beta H_0}e^{ixX}\,A_{j}^{\dag}(it).
\end{equation}
(Here, we use imaginary time from the start.) When calculating $\Tr\overline{W}^s$ for varying $s$, it is convenient to replace $\beta$ with $\beta/s$. Thus,
\begin{equation}
\frac{\partial_\epsilon\Tr(\overline{W}^s)|_{\epsilon=0}}{s}
=\sum_j\Tr\bigl(e^{-\frac{(s-1)\beta H_0}{s}}e^{ixX}
A_j(it)e^{-ixX}e^{-\frac{\beta H_0}{s}}e^{ixX}A^\dagger_j(it)e^{-ixX}\bigr).
\end{equation}
Notice the cancellation of $e^{ixX}$ and $e^{-ixX}$ between different replicas. To the order of $x^2$, there are only four OTOCs, where $A$ and $A^\dag$ belong to one replica, one copy of $X$ is sandwiched between them, and another $X$ belongs to an adjacent replica:
\begin{equation}
\Tr\bigl(e^{-\frac{(s-1)\beta H_0}{s}}XA_j(it)X
e^{-\frac{\beta H_0}{s}}A^\dagger_j(it)\bigr)
-\Tr\bigl(e^{-\frac{(s-1)\beta H_0}{s}}A_j(it)X
e^{-\frac{\beta H_0}{s}}A^\dagger_j(it)X\bigr) +\text{h.c.}
\end{equation}
This can be written as
\begin{equation}
\langle A_j^\dagger(\beta+it)X(\beta/s)A_j(\beta/s+it)X(\beta/s)\rangle
-\langle A_j^\dagger(\beta+it)X(\beta)A_j(\beta/s+it)X(\beta/s)\rangle
+\text{h.c.}
\end{equation}
Applying the single-mode ansatz to these OTOCs, we find the entropy change to be expressed as an integral similar to~(\ref{BB1}--\ref{BB2}) with
\begin{equation}
f_S(s,\omega_{14},\omega_{23})=e^{-\frac{(s-1)\beta \omega_{14}}{s}}\left(2\cos\left(\frac{-1+2/s}{4}\varkappa \beta\right)-2\cos\left(\frac{\varkappa \beta}{4}\right)e^{-\frac{(s-1)\beta \omega_{23}}{s}}\right).
\end{equation}
Taking the derivative with respect to $s$, we obtain the final expression:
\begin{equation}
f_S(\omega_{14},\omega_{23})=\partial_s f_S(s,\omega_{14},\omega_{23})|_{s=1}=2\beta \omega_{23}\cos\left(\frac{\varkappa \beta}{4}\right)+\beta \varkappa\sin\left(\frac{\varkappa \beta}{4}\right).
\end{equation}

\section{Summary and discussion}\label{sec:discussion}

We have computed the von Neumann entropy in certain many-body settings by combining time-dependent perturbation theory and the replica trick. In particular, we have considered a system made of two parts that are prepared in the thermofield double state, and then one part is coupled to a heat bath. The von Neumann entropy of the two double system grows as $S(\rho(t))\approx at\ln t^{-1}$ at very short times and as $S(\rho(t))\approx bt$ at longer times (but less than the Page time); the coefficients $a$ and $b$ have been calculated.
These calculations involve unusual terms, proportional to $\kappa^{2s}$, in the perturbative expansion in the system-bath coupling strength $\kappa$, where $s$ is the number of replicas.

In the second part of the paper, we developed a variant of thermodynamic response theory. In general, thermodynamic response is concerned with perturbing the system with operators $X_1(\tau_1),\dots,X_n(\tau_n)$ taken with coefficients $x_1,\dots,x_n$. The standard theory uses the generating function $\ln Z$, whereas our modified version is based on $\ln Z-S$. Thus, the connected correlator turns into a so-called branched correlator,
\begin{equation}
\corr{\BB,X_n(\tau_n),\cdots,X_1(\tau_1)}
= \left.\frac{\partial^n}{\partial x_1\cdots\partial x_n}\right|
_{x_1=\dots=x_n=0}\Bigl(\ln Z-S\Bigr).
\end{equation}
($\BB$ is not an independently defined object but rather, part of the notation.) While the calculation of branched correlators is rather involved, a simplification occurs for maximally chaotic systems. We now argue that this special case is consistent with a holographic picture in a very broad sense.

For comparison, consider a Euclidean black hole in a hyperbolic space (say, in two dimensions). The replica geometry involves an $s$-fold cover of both the circle and the disk it bounds, with a branching point at the center. Inserting boundary fields slightly deforms the space. In the $s\to 1$ limit, the geometry is given by a smooth metric on the disk and the position of the branching point. The branching point is a special case of a quantum extremal surface~\cite{EnWa14} (where ``surface'' means a codimension $2$ submanifold). Its position is determined by an extremum of entropy. Instead of the entropy $S$, we may consider $\ln Z-S$. Indeed, the partition function $Z$ depends only on the space-time metric, which should be fixed before finding the extremal surface.

In the Lorentzian case, the branching point is described by null coordinates $(u_{+},u_{-})$. (We set aside the ambiguity in the choice of origin due to the deformation of space-time relative to $\operatorname{AdS}_2$.) The entropy can be expanded to the second order in $u_{+}$, $u_{-}$:
\begin{equation}
S(u_{+},u_{-})=S_0+p_{+}u_{+}+p_{-}u_{-}-Cu_{+}u_{-},
\end{equation}
where $p_{+}$ and $p_{-}$ depend on the inserted field.\footnote{This expression is similar to 't~Hooft's effective action~\cite{tH90} for the fluctuating horizon, where $p_{+}$ and $p_{-}$ are null energies. In our case, they are just abstract coefficients.} Solving the extremum problem, we get
\begin{equation}
S_{\mathrm{ext}}=S_0+C^{-1}p_{+}p_{-}.
\end{equation}

Now, let us forget about geometry. The only property we need is that if there is large time separation, then $p_{+}$ and $p_{-}$ depend only on the fields inserted in the past and the future, respectively. Thus, the change in the entropy should factor into two quantities dependent on the corresponding fields. This is exactly what we observed in the maximal chaos case, see equation \eqref{f1423_simple}. We leave the interpretation of these quantities to future research.

\section{Acknowledgments}

We thank Douglas Stanford, Juan Maldacena, and Yiming Chen for useful discussions. We gratefully acknowledge the support by the Simons Foundation under grant~376205. A.K.\ is also supported by the Simons Foundation through the ``It from Qubit'' program, as well as by the Institute of Quantum Information and Matter, a NSF Frontier center funded in part by the Gordon and Betty Moore Foundation. P.Z.\ acknowledges support from the Walter Burke Institute for Theoretical Physics at Caltech.

\bibliography{PertEntropy}
\bibliographystyle{JHEP}

\end{document}